\documentclass[%
 reprint,
 superscriptaddress,
 amsmath,amssymb,
 aps,
 prb,
]{revtex4-2}

\usepackage{graphicx}% Include figure files
\usepackage{dcolumn}% Align table columns on decimal point
\usepackage{bm}% bold math
\usepackage{dsfont}% special shape latters
\usepackage{braket}% Dirac notation
\usepackage{hyperref}% add hypertext capabilities
\usepackage{xcolor}
\usepackage{soul}
\usepackage{changes}
\usepackage{ulem}
\usepackage[utf8]{inputenc}
\usepackage{cases}

\newcommand{\RNum}[1]{\uppercase\expandafter{\romannumeral #1\relax}}% for Roman number
\let\vec\mathbf % vector as bold form

\UseRawInputEncoding
\begin{document}

\author{Zhihao Jiang}
\author{Stephan Haas}
\affiliation{Department of Physics and Astronomy, University of Southern California, Los Angeles, California 90089-0484, USA}

\author{Malte R\"{o}sner} % corresponding author
\email{M.Roesner@science.ru.nl}
\affiliation{Institute for Molecules and Materials, Radboud University, Heyendaalseweg  135, 6525 AJ Nijmegen, The Netherlands}

\date{\today}

\title{Plasmonic Waveguides from Coulomb-Engineered Two-Dimensional Metals}

\begin{abstract}
  
  Coulomb interactions play an essential role in atomically-thin materials. On one hand, they are strong and long-ranged in layered systems due to the lack of environmental screening. On the other hand, they can be efficiently tuned by means of surrounding dielectric materials. Thus all physical properties which decisively depend on the exact structure of the electronic interactions can be in principle efficiently controlled and manipulated from the outside via Coulomb engineering. Here, we show how this concept can be used to create fundamentally new plasmonic waveguides in metallic layered materials. We discuss in detail how  dielectrically structured environments can be utilized to non-invasively confine plasmonic excitations in an otherwise homogeneous metallic 2D system by modification of its many-body interactions. We define optimal energy ranges for this mechanism and demonstrate plasmonic confinement within several nanometers. In contrast to conventional functionalization mechanisms, this scheme relies on a purely many-body concept and does not involve any direct modifications to the active material itself.
  
\end{abstract}
\maketitle

\section{Introduction}

Plasmons are collective excitations rendered by  dynamical screened Coulomb interactions. They are hence intimately connected to a plethora of fundamental many-body material properties of electronic systems, such as 
their quasi-particle spectral function \cite{Hedin1965, Aryasetiawan1998, Onida2002,  Kas2014},
optical absorption spectra \cite{Onida2002, Steinhoff2017},
 light-matter interactions \cite{Barnes2003, Low2017}, and possibly also to instabilities such as superconductivity \cite{Bill2003, Hepting2018}, charge-density order \cite{VanWezel2011}, or excitonic condensation \cite{Kogar2017}. 
From a technological point of view, plasmonic structures such as  antennas and waveguides have become increasingly important tools to create efficient light harvesting and guiding devices \cite{Oulton2008,Ansell2015,Fang2015,Ditlbacher2005,Bozhevolnyi2006,Barnes2003,Schnell2011,Boltasseva2008,Kress2015}. 
In this context, layered materials are of particular interest due to their enhanced low-energy plasmonic response. In contrast to three-dimensional bulk materials, two-dimensionally confined plasmons have a gapless excitation spectrum with a square-root dispersion at small momenta \cite{DaJornada2020, Groenewald2016, Andersen2013, Liu2008}. As  will be discussed in detail below, this yields extended excitation energy ranges in which low losses are expected \cite{Gjerding2017, Woessner2015}, and  which can be precisely controlled from the outside by Coulomb engineering, thus rendering 2D plasmonics an exciting and technologically important field. 

\begin{figure}[htbp]
 \centering
 \includegraphics[width=0.4\textwidth]{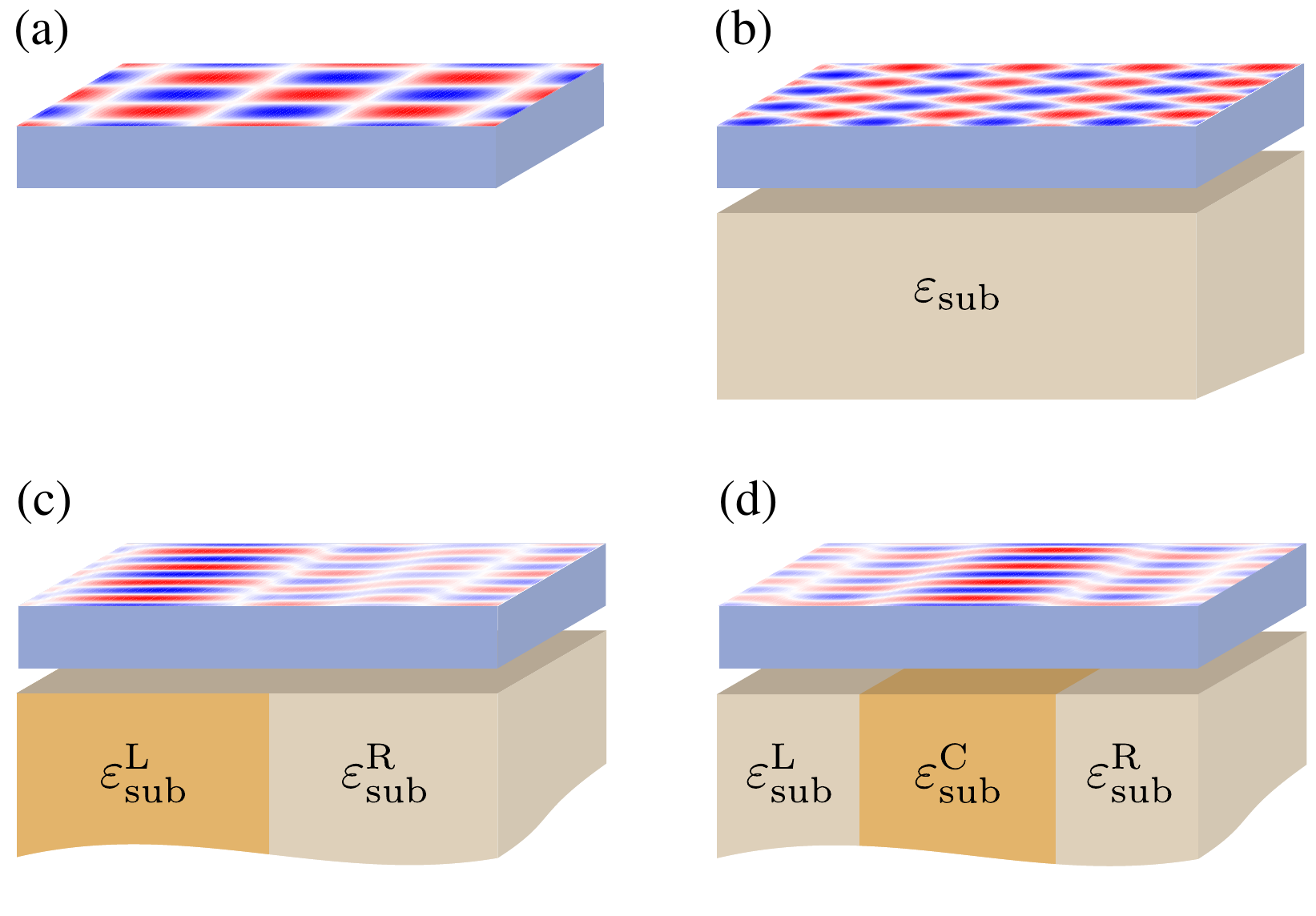}
 \caption{\textbf{Substrate controlled plasmon patterns in atomically thin metals.} (a) Unsupported and (b)-(d) dielectrically supported metallic layer with a typical plasmonic charge distribution. (c),(d) Structured dielectric substrates induce spatially patterned plasmonic excitations. }\label{fig:PWG_sketch}
\end{figure}

So far, two-dimensional (2D) plasmonic devices have been manufactured either by strongly invasive processes, such as structuring the activate material itself \cite{Nikitin2011,Yuan2013,Christensen2012,Kim2011,Thongrattanasiri2012}, by changes to the single-particle properties of the plasmon-hosting system \cite{Vakil2011,Liu2011,Prishchenko2018,Slotman2018}, or by creating metallic heterostructures \cite{Iranzo2018}.
Here, we propose a truly non-invasive concept based on the external control of the Coulomb interactions within the active material.  Specifically, we show how 2D plasmons in an homogeneous layered material can be spatially controlled with the help of structured dielectric environments, as depicted in Fig.~\ref{fig:PWG_sketch}. An analogous Coulomb engineering concept has been previously applied to atomically thin semiconductors \cite{Rosner2016,Steinke2020,Raja2017,Florian2018,Steinleitner2018,VanLoon2020,Utama2019,Kajino2019}.
Here, we apply this concept to active metallic systems. Specifically, we will consider homogeneous as well as horizontally staggered dielectric environments. By realistically accounting for horizontal and vertical dielectric-interface effects via image charge models within a real space random phase approximation, we show that plasmons can be precisely confined at certain optimal excitation energies. This allows for the non-invasive creation of plasmonic waveguides, as illustrated in Fig.~\ref{fig:PWG_sketch}\,(d), whereby the active, plasmon hosting, monolayer remains unchanged. 

\section{Results and Discussion}
\subsection{Substrate Controlled 2D Plasmons}

\begin{figure}[htbp]
 \centering
 \includegraphics[width=8.6cm]{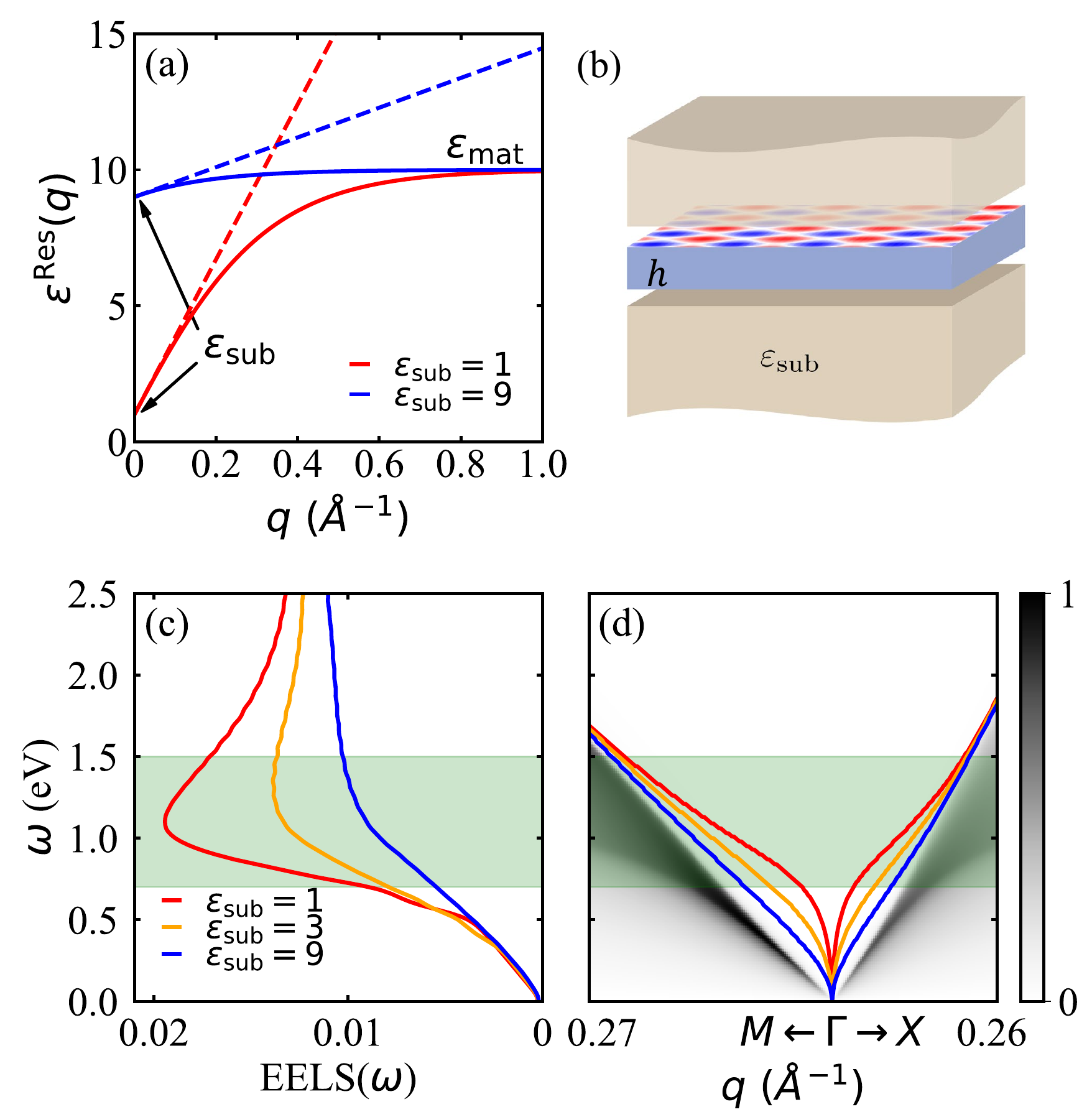}
 \caption{\textbf{Substrate controlled two-dimensional plasmons.} (a)  Effective residual screening function $\varepsilon^\text{Res}(q)$ for different homogeneous dielectric substrates. The dashed lines indicate the slopes of the function at $q=0$. (b) Illustration of plasmon excitation in a layered material embedded in a homogeneous dielectric environment. (c) Electron energy loss spectra of the two-dimensional metallic monolayer shown in (b), with different choices of the environmental dielectric constant. (d) Plasmon dispersions in  momentum space around $\Gamma$. The grey area represents the electron-hole continuum.}\label{fig2_homo_qspace}
\end{figure}

In the following, we describe the plasmonic excitations in layered metals using a single-orbital generalized Hubbard model of the form
\begin{align}
    \hat{H} =  
        - t \sum_{\sigma,<ij>} &\hat{c}_{i\sigma}^{\dagger}\hat{c}_{j\sigma} \\
        + \sum_{\sigma,i} \ &U_{ii} n_{i\sigma}n_{i\bar{\sigma}} 
        + \sum_{\sigma,\sigma',i>j}U_{ij} n_{i\sigma}n_{j\sigma'} \notag, \label{2D_Hamiltonian}
\end{align}
where $\hat{c}_{i\sigma}^{\dagger}$ ($\hat{c}_{i\sigma}$) creates (annihilates) an electron of spin $\sigma$ at the square-lattice site $i$, $n_{i\sigma}$ are the corresponding occupation operators, $t=1\,$eV is the nearest neighbour hopping, and $U_{ij}$ is the non-local Coulomb interaction. We fix the Fermi energy to $E_F=-3\ \text{eV}$ and treat the model within the random phase approximation. 
To realistically describe the collective excitations of a generic layered metal in terms of this model we need to account for all involved screening channels. This includes intra- and inter-band polarizations within the layered material itself \cite{Andersen2013,Groenewald2016,DaJornada2020}, as well as external polarization effects from the environment, such as substrates or capping layers. To this end we consider the fully screened Coulomb interaction in the metallic film  for in-plane momentum transfer $q$,
\begin{align}
    W(q, \omega) = \frac{v_q}{1 - v_q \Pi(q,\omega)} = \frac{v_q}{\varepsilon(q,\omega)},
\end{align}
where $\Pi(q,\omega)$, $\varepsilon(q,\omega)$, and $v_q = \frac{2\pi e^2}{q}$ are the total polarization, the dielectric function, and the bare Coulomb interaction, respectively. 
By splitting $\Pi = \Pi^\text{M} + \Pi^\text{Res}$ into the metallic polarization, resulting from intra-band transitions within the active layered material itself described by the Hamiltonian above, and a residual polarization, accounting for inter-band transitions and substrate screening, we can rewrite the fully screened Coulomb interaction as
\begin{align}
    W(q, \omega) = \frac{U_q}{1 - U_q \Pi^\text{M}(q,\omega)} = \frac{U_q}{\varepsilon^\text{M}(q, \omega)},
\end{align}
where $U_q = \frac{v(q)}{\varepsilon^\text{Res}(q)}$ is the Fourier transform of $U_{ij}$ accounting for the residual screening channels. 
The full dielectric function is thus given by the product $\varepsilon(q,\omega) = \varepsilon^\text{M}(q, \omega) \varepsilon^\text{Res}(q)$. 
We calculate $\varepsilon^\text{M}(q, \omega)$ within the Random Phase Approximation (RPA) based on the Hamilton from above and by neglecting local-field effects, while the residual screening can be analytically approximated from classical electrostatics, which reads for a layered material of effective height $h$ and intrinsic dielectric constant $\varepsilon_\text{mat}$ \cite{Andersen2015,Cho2018,Groenewald2016,Rosner2016}
\begin{align}
    \varepsilon^\text{Res}(q) = \varepsilon_\text{mat}\frac{1-\tilde{\varepsilon}^2e^{-2qh}}{1+2\tilde{\varepsilon}e^{-qh}+\tilde{\varepsilon}^2e^{-2qh}}
\end{align}
For a dielectric encapsulation, as shown in  Fig.~\ref{fig2_homo_qspace}~(b), we  find $\tilde{\varepsilon}=(\varepsilon_\text{mat}-\varepsilon_\text{sub})/(\varepsilon_\text{mat}+\varepsilon_\text{sub})$.
This allows us to define the electron energy loss spectrum (EELS),
\begin{align}
    \text{EELS}(q,\omega) 
        \propto - \operatorname{Im} \left[ \frac{1}{\varepsilon(q,\omega)} \right].
\end{align}
According to the implicit definition of plasmonic excitations, $\varepsilon(q,\omega_p(q)) = 0$, the electron energy loss spectrum is maximized along the plasmonic dispersion $\omega_p(q)$. This way, we can extract $\omega_p(q)$ along a path in momentum space, shown in Fig.~\ref{fig2_homo_qspace}~(d), along with the metallic polarization function $\Pi^\text{M}(q,\omega)$. 
The model parameters are chosen to approximately reproduce the plasmonic energy scales of metallic transition metal dichalcogenides \cite{Groenewald2016, Andersen2015, Andersen2013} and doped hexagonal boron nitride \cite{Loncaric2018}. 
We observe that even in the free-standing case ($\varepsilon_\text{sub} = 1$) the plasmonic dispersion deviates quickly from the generic $\sqrt{q}$-like dispersion \cite{Andersen2013,Groenewald2016,DaJornada2020}, known from purely two-dimensional models. This deviation is a consequence of the non-local residual screening  $\varepsilon^\text{Res}(q)$ induced by inter-band transitions within the layered metal, which we plot in Fig.~\ref{fig2_homo_qspace}~(a).
For the free-standing case, its long and short wavelength limits are $\varepsilon^\text{Res}(q=0) = 1$ and $\varepsilon^\text{Res}(q \gg 1) = \varepsilon_\text{mat}$, respectively. Hence, in the long wavelength limit the effective background screening is negligible \cite{Lundeberg2017}, and $\omega_p(q) \propto \sqrt{q}$ holds. For larger momenta, however, $\varepsilon^\text{Res}$ increases successively, which suppresses $\omega_p(q)$ and eventually pushes it into the particle-hole continuum. Importantly, the corresponding flattening of the plasmonic dispersion takes place at momenta which are clearly detached from the particle-hole continuum, so that Landau damping and thus plasmonic losses are drastically reduced in layered metals at small and intermediate momenta \cite{DaJornada2020,Loncaric2018,Andersen2013,Groenewald2016}. This, in turn, leads to a prominent enhancement of the plasmonic spectral function at intermediate frequencies, as it is clearly visible in the $\text{EELS}(\omega)$, shown in Fig.~\ref{fig2_homo_qspace}~(c) (green shaded region). 
Upon increasing the environmental screening, e.g. by using different substrate materials [illustrated in Fig.~\ref{fig2_homo_qspace}~(b)], the long wavelength limit of the effective residual screening is changed to $\varepsilon^\text{Res}(q=0) = \varepsilon_\text{sub}$. This leads to a decreased $U_q$, and subsequently to a reduced plasmonic dispersion with enhanced slopes for small momenta. In the full $\text{EELS}(\omega)$ we correspondingly find a decreased and broadened maximum at intermediate frequencies, whereas the remainder of the plasmonic spectrum is largely unaffected. This intermediate frequency range in the plasmonic spectral function is thus most susceptible to changes in the environmental screening of the layered material. 

\begin{figure}[htbp]
 \centering
 \includegraphics[width=8.6cm]{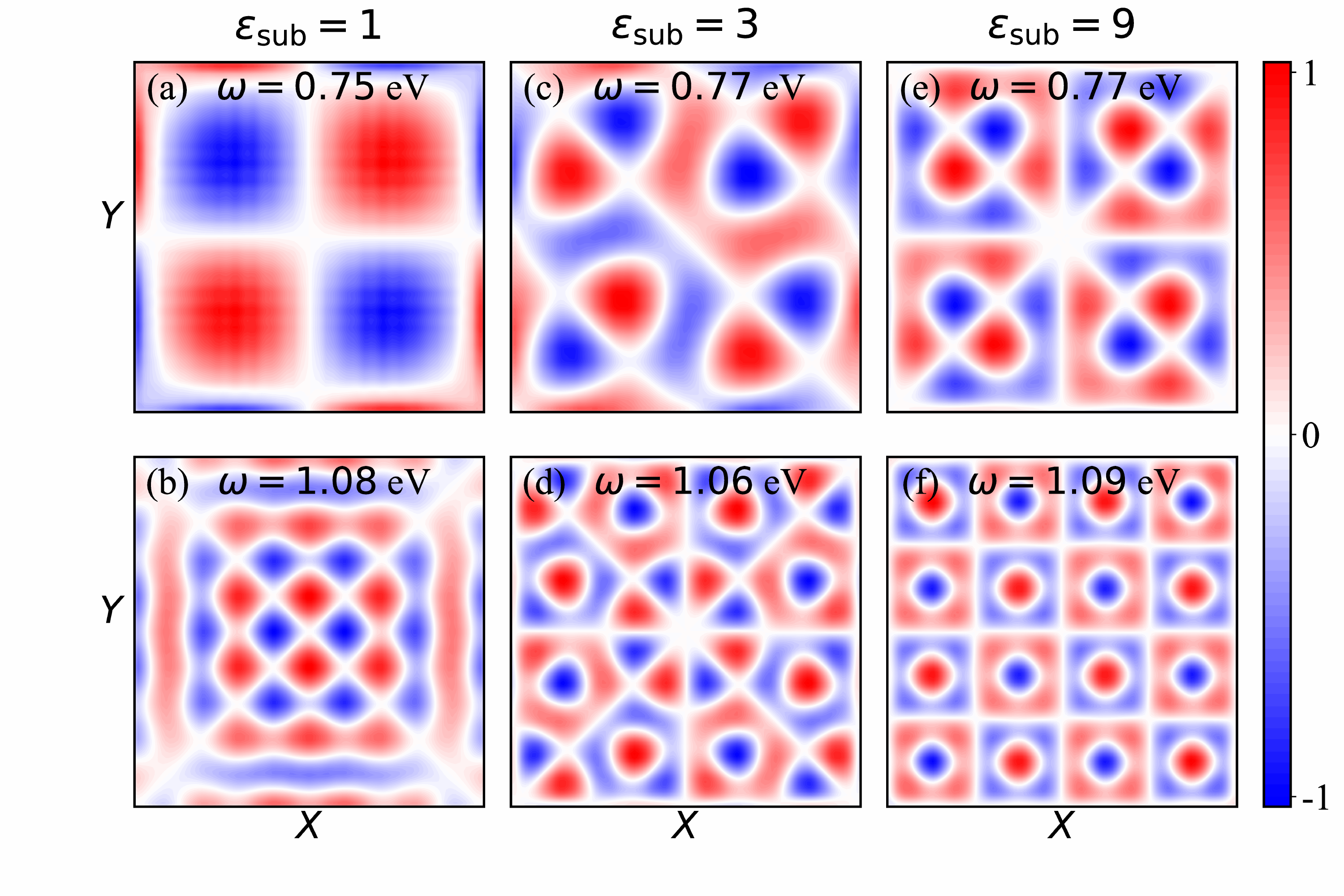}
 \caption{\textbf{Real-space charge density modulations of two typical plasmon modes.} At $\omega \approx 0.77\ \mathrm{eV}$ and $\omega \approx 1.08\ \mathrm{eV}$, plotted for three different dielectric environments.}\label{fig3_homo_modes}
\end{figure}

The enhanced environmental screening sensitivity of plasmons in layered materials at intermediate excitation energies is also reflected in the corresponding real space charge density patterns, shown in Figs.~\ref{fig3_homo_modes}~(a-f) at $\omega \approx 0.77$eV and  $\omega \approx 1.08\,$eV for the same $\varepsilon_\text{sub}$ as before. Here, we observe extended modes with checkerboard-like patterns oscillating along the $x$ and $y$ directions. At a fixed excitation energy, these patterns show decreasing wavelengths upon increasing the environmental screening, in line with the corresponding dispersions shown in Fig.~\ref{fig2_homo_qspace}~(d). Similarly, we observe that the plasmonic wavelengths decrease with increasing excitation energies at fixed screening.

This demonstrates how spatial patterns of 2D plasmons can be  controlled by means of dielectric substrates. In the following, we illustrate how this can be utilized to spatially confine plasmonic excitations by using structured dielectric environments.

\subsection{2D Plasmons in Heterogeneous Screening Environments}

\begin{figure}[htbp]
 \centering
 \includegraphics[width=8.6cm]{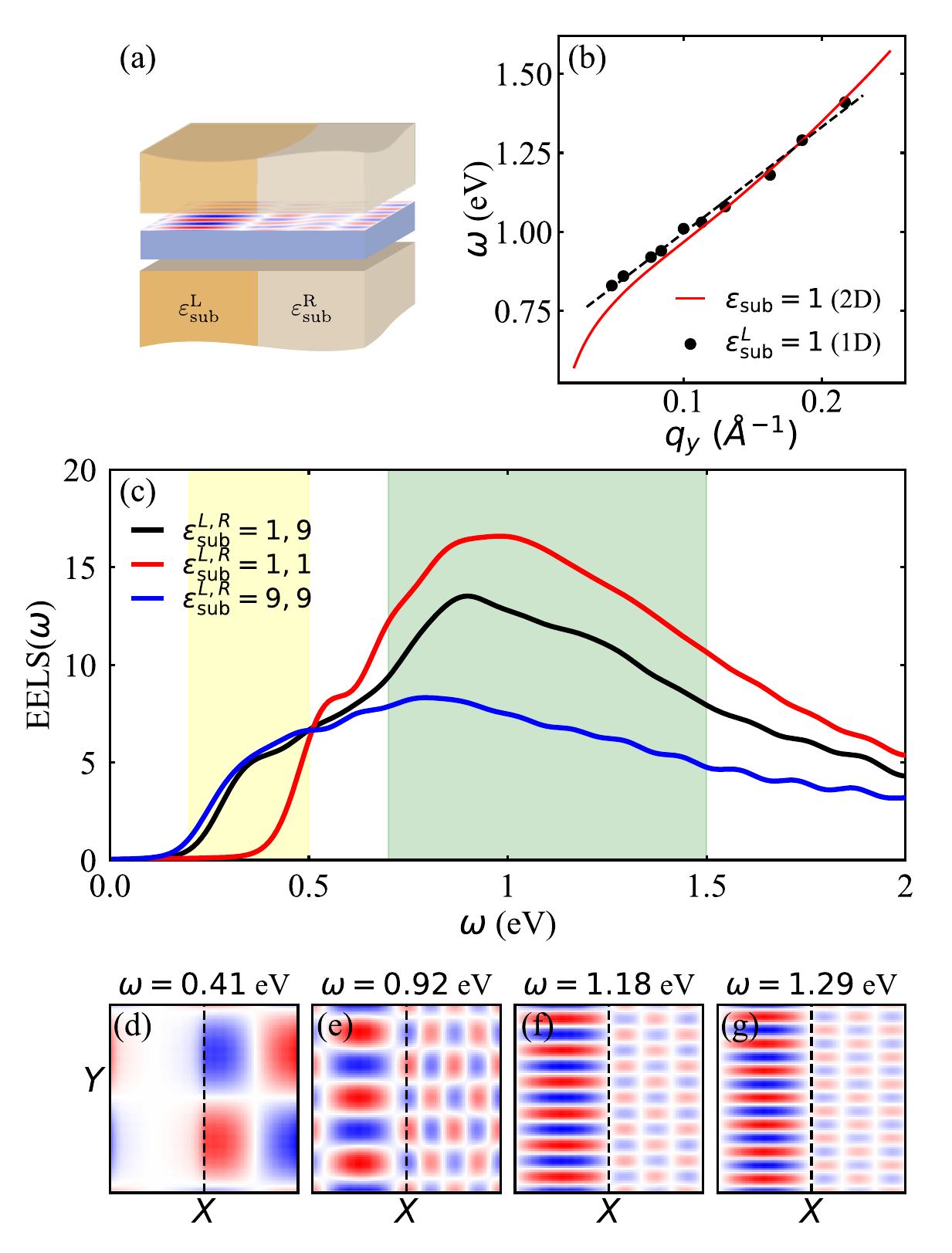}
 \caption{\textbf{Heterogeneous plasmon patterns from spatially structured dielectric environments.} (a) Sketch of the layered metal embedded in a heterogeous dielectric environment. (b) Dispersion relation of the plasmon mode in the left subspace with $\varepsilon_\text{sub}^\text{L}=1.0$ and $\varepsilon_{\text{sub}}^{\text{R}}=9$ (black dots) together with the dispersion from Fig.~\ref{fig2_homo_qspace}~(c) for $\varepsilon_{\text{sub}} = 1$.  (c) Total EELS (black) for $\varepsilon_{\text{sub}}^{\text{L}}=1$ and $\varepsilon_{\text{sub}}^{\text{R}}=9$ together with homogeneous EELS (red and blue). (d-g) Typical plasmon modes at different frequencies.}\label{fig4_hetero_19}
\end{figure}

Let us now consider a heterogeneous dielectric environment with a vertical interface separating areas with $\varepsilon_\text{sub}^\text{L}=1$ and $\varepsilon_\text{sub}^\text{R}=9$, as depicted in Fig.~\ref{fig4_hetero_19}~(a). Since this breaks the translational symmetry along the $x$-direction, we are forced to switch to a real space representation. To this end, we utilize a supercell consisting of $80 \times 80$ unit cells. In Fig.~\ref{fig4_hetero_19}~(c) we show the resulting heterogeneous EELS$(\omega)$ (black) as well as the corresponding EELS$(\omega)$ for the two homogeneous situations (red and blue). By comparing the heterogeneous spectrum to the unsupported one (i.e. $\varepsilon_\text{sub}^\text{L} = \varepsilon_\text{sub}^\text{R} = 1$), we observe that the maximum at intermediate excitation energies (green shaded) is still present, but suppressed. Furthermore, we see that at low energies (yellow shaded) a new shoulder arises, which aligns with the onset of the spectral function of a homogeneously supported monolayer with $\varepsilon_\text{sub}^\text{L} = \varepsilon_\text{sub}^\text{R} = 9$. The heterogeneous spectral function thus clearly inherits characteristics from both homogeneous limits. The origin of this becomes obvious by examining the corresponding real space charge modulations. At small excitation energies, the plasmons are mostly localized in the $\varepsilon_\text{sub}^\text{R}$ region, as seen in Fig.~\ref{fig4_hetero_19}~(d), whereas at intermediate excitation energies they prominently reside in the $\varepsilon_\text{sub}^\text{L}$ area, as observed in Figs.~\ref{fig4_hetero_19}~(e-g).

This illustrates how heterogeneous plasmonic patterns can be externally and non-invasively  induced in homogeneous layered materials via spatially structured substrates. The substrate induced heterogeneous plasmonic patterns behave, however, slightly differently compared to their homogeneous counterparts depicted in Fig.~\ref{fig3_homo_modes}. For intermediate excitation energies, which mostly confine the plasmon in the $\varepsilon_\text{sub}^\text{L}$ region, the resulting pattern is now quasi-one-dimensional. Specifically, we observe that its propagation direction aligns with the dielectric interface in the substrate and has a \textit{linear} dispersion, as shown in Fig.~\ref{fig4_hetero_19}~(b).
Additionally, there are some spurious, strongly damped, plasmonic excitations present in the $\varepsilon_\text{sub}^\text{R}$ region.

Interestingly, the fully heterogeneous EELS [black line in Fig.~\ref{fig4_hetero_19}~(c)] can be approximately reconstructed by taking a simple average over the homogeneous data (blue and red lines). From this we understand that the $\varepsilon_\text{sub}^\text{L}=1$ environement barely affects the low-energy plasmonic excitations confined in the $\varepsilon_\text{sub}^\text{R}=9$ area, which does not hold vice versa. I.e. the $\varepsilon_\text{sub}^\text{R}=9$ slightly damps the plasmonic excitations on the $\varepsilon_\text{sub}^\text{L}=1$ side. However, except from this, these two patterns behave largely independently on the other side.

\begin{figure}[htbp]
 \centering
 \includegraphics[width=8.6cm]{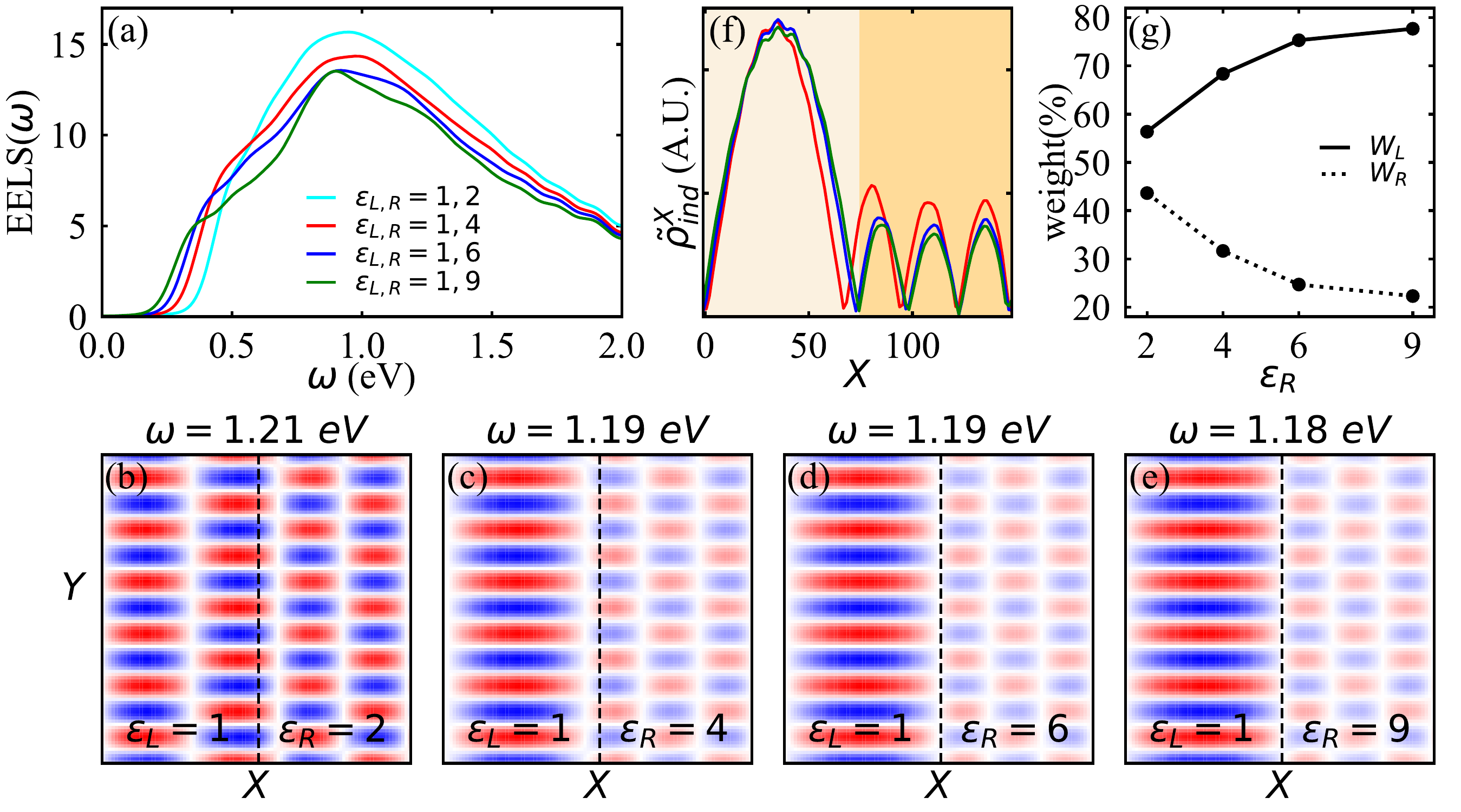}
 \caption{\textbf{Impact of the dielectric contrast to the plasmonic patterns.} (a) Total EELS for a different dielectric contrasts. (b-e) Corresponding real-space patterns for $\omega \approx 1.2\,$eV. (f) Normalized $x-$component of the charge distributions of these plasmon modes. (g) Variation of the charge weight in the left ($W_L$) and right ($W_R$) subspaces. }\label{fig5_hetero_contrast}
\end{figure}

We proceed with a quantitative analysis of these confined excitations by investigating their dependence on the dielectric contrast ratio $\varepsilon_\text{sub}^\text{R}/\varepsilon_\text{sub}^\text{L}$ in the substrate. To this end, we fix $\varepsilon_\text{sub}^\text{L}=1$ and vary $\varepsilon_\text{sub}^\text{R}$. 
In Fig.~\ref{fig5_hetero_contrast}~(a) we observe that increasing the contrast affects the EELS in a non-trivial way.
At small dielectric contrasts, the full heterogeneous EELS resembles a homogeneous one, with just one broad maximum at intermediate excitation energies, which is reflected in the rather spread-out plasmonic excitation pattern depicted in Fig.~\ref{fig5_hetero_contrast}~(b). Upon increasing the substrate dielectric contrast, we find a low-energy energy shoulder arising and shifting to lower frequencies. 
However, overall we find an increasing localization of the plasmonic pattern on the $\varepsilon_\text{sub}^\text{L}$ side, with decreasing weight on the $\varepsilon_\text{sub}^\text{R}$ side for $\omega \approx 1.2\,$eV, as depicted in Figs.~\ref{fig5_hetero_contrast}~(c-e). We can furthermore quantify this increasing localization with the help of the normalized charge distribution function, $\tilde{\rho}_{ind}(x)$,  obtained by integrating the absolute value of the real space charge distribution function along the $y$-component [see Fig.~\ref{fig5_hetero_contrast}~(f)]. By additionally integrating $\tilde{\rho}_{ind}(x)$ over the left and right regions we can define sub-space weights $W_\text{L/R}$, which we show in Fig.~\ref{fig5_hetero_contrast}~(g). Then we observe that up to nearly $80\%$ of the charge can be confined within the left region for the maximum dielectric contrast considered here. As a result the plasmonic excitation in the $\varepsilon_\text{sub}^\text{L}$ area gets relatively brighter. However, increasing $\varepsilon_\text{sub}^\text{R}$ also damps the excitation in the $\varepsilon_\text{sub}^\text{L}$ region. Therefore, achieving an optimal dielectric contrast will be a trade-off between spatial contrast and brightness of the plasmons on the active side. Finally, we note that the $\varepsilon_\text{sub}^\text{L}$-confined quasi-one-dimensional plasmon wavelength is not affected by $\varepsilon_\text{sub}^\text{R}$.

\subsection{Novel Plasmonic Waveguides}

\begin{figure}[htbp]
 \includegraphics[width=8.6cm]{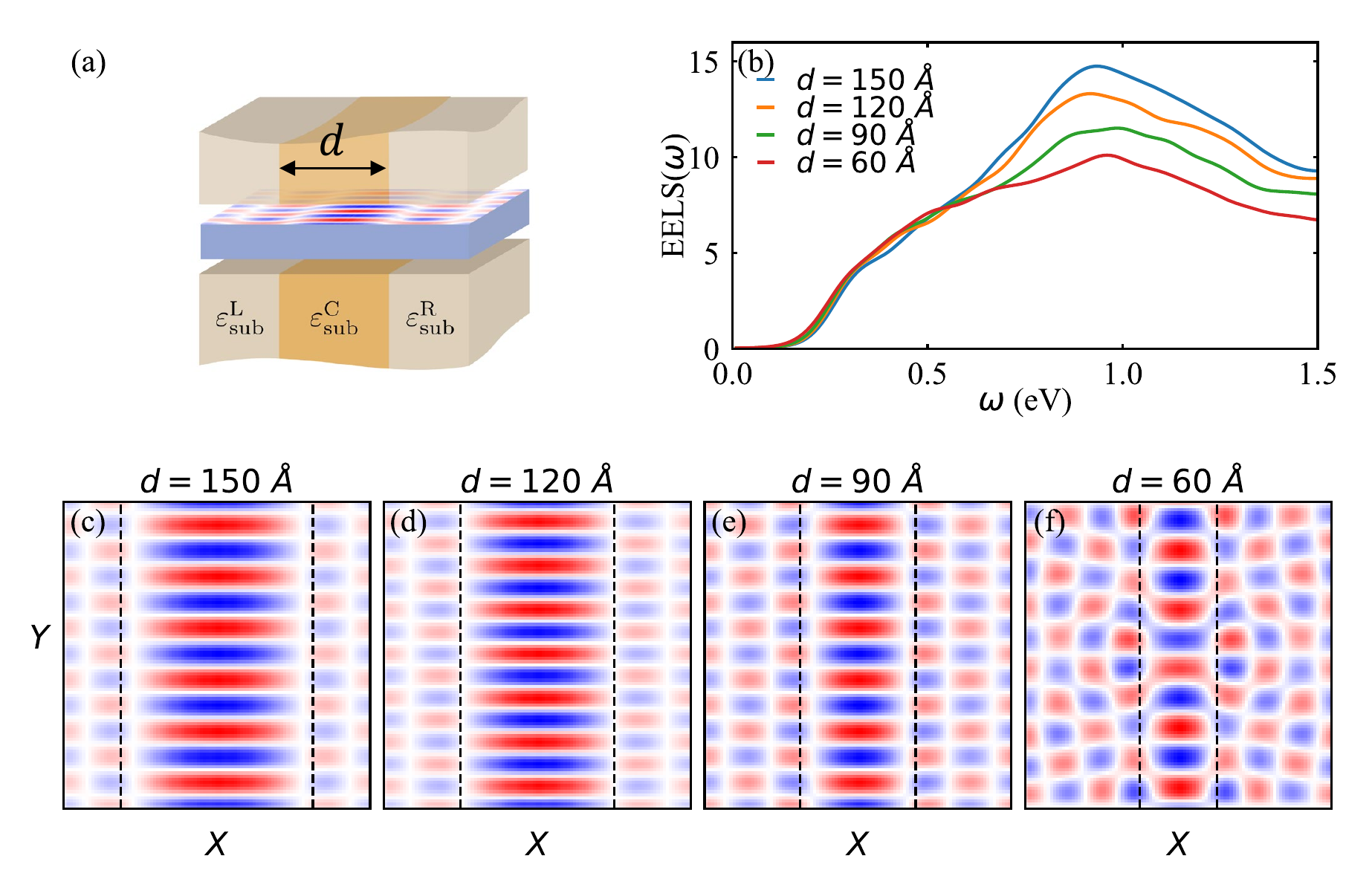}
 \caption{\textbf{Plasmonic waveguides from spatially structured dielectric environments.} (a) Illustration of the plasmonic waveguide design. (b) Dependence of the EELS on the core width $d$. (c-f) Real space charge modulations of typical plasmon modes at $\omega \approx 1.0\,$eV.}\label{WG}
\end{figure}

The possibility to non-invasively spatially pattern plasmonic excitations in a homogeneous layered material by means of dielectric interfaces in the dielectric environment motivates us to propose a new class of plasmonic waveguides that utilizes two parallel vertical dielectric interfaces in the environment, as illustrated in Fig.~\ref{WG}~(a). This will confine well-defined quasi-one-dimensional plasmons in the central region if the substrate's dielectric constant there is smaller than in the outer substrate regions. To verify this proposal, we study below a system with $\varepsilon_\text{sub}^\text{L} = \varepsilon_\text{sub}^\text{R} = 9$ and $\varepsilon_\text{sub}^\text{C} = 1$, and with variable central substrate width $d$.

In Fig.~\ref{WG}~(a) we show the EELS for various $d$. Starting from large $d$, we recover the previously discussed maximum at intermediate excitation energies which is accompanied by a low-energy shoulder. Upon decreasing $d$, this maximum diminishes until it vanishes below $d \lesssim 60\,$\AA\ ($20$ unit cells). This behavior becomes clear by examining the corresponding real space patterns shown in Figs.~\ref{WG}~(c-f). Fig.~\ref{WG}~(c) reveals that the plasmon mode is now indeed spatially confined in the central substrate area and propagates only parallel to the substrate dielectric interfaces. Thus, the maximum in EELS$(\omega)$ results from the low-dielectric substrate region, and the correspondingly confined plasmonic excitation there. Upon decreasing $d$, we observe in Figs.~\ref{WG}~(d-f) that the novel plasmonic waveguide behavior persists, but with gradually decreasing contrast to the $\varepsilon_\text{sub}^\text{L/R}$ areas, until it nearly vanishes for the smallest $d$ shown in Fig.~\ref{WG}~(f). Thus, the increasing environmental screening from the increasing $\varepsilon_\text{sub}^\text{L/R}$ regions gradually damps the confined excitation in the center of the device until no spatial structure is visible anymore. Nevertheless, a clear confinement can be achieved for waveguides with widths down to about $30$ unit cells, which is on the order of $90\,$\AA\ here. 
Like in the case of the dielectric contrast, there is a  trade-off between the field confinement and loss in the optimization of these plasmonic waveguides \cite{Kress2015,Tame2013}. 
This is a demonstration of a novel type of plasmonic waveguides,  based on Coulomb-engineered  homogeneous layered metallic materials.

\section{Outlook and Conclusions}

    \textit{Active Material Candidates:} Apart from  the properties of the structured dielectric environment, our proposal also depends strongly on the active material itself, i.e. not all 2D metals will be equally suitable. Specifically, the active metallic layer should be very sensitive to environmental screening. This is the case if all layer-internal polarization channels (intra- and inter-band) are rather small, in other words: materials with a small density of states at the Fermi level and with a metallic (low-energy plasmon hosting) band which is energetically well separated from all other valence and conduction bands. Both of these properties are satisfied in graphene: it hosts low-loss plasmons \cite{Jablan2009,Low2014,Christensen2012,Grigorenko2012,Luo2013}, and the Coulomb interactions have been experimentally shown to be rather susceptible to external polarization \cite{Lundeberg2017,Kim2020,Iranzo2018}.
    Alternatively, slightly doped semi-conducting layered materials, such as electron or hole doped MoS$_2$ or WS$_2$, could be suitable candidates due to reduced intra-band polarization as well as metallic $3R$-NbS$_2$ and $1T$-AlCl$_2$ due to their low internal plasmonic losses \cite{Andersen2015}.
    
    \textit{Structured Environments: } For the creation of structured environments, as needed for the proposed waveguides,  we envision  laterally grown and vertically cut lithographic structures, twisted layered materials \cite{Bistritzer2011,Andrei2020,Weston2020}, or novel fractionalized 2D system \cite{Westerhout2018,He2019,DeNicola2020} to be possible routes to pursue. As discussed above, the dielectric contrast in these structures should be rather high. Furthermore, it will be interesting to create environmental screening structures with more than just one or two dielectric interfaces. One could, for example, imagine periodically patterned substrates for optimal light-matter coupling, two-dimensional dielectric structures that either create plasmonic checkerboard patterns or plasmonic quantum dots, or non-linear plasmonic waveguides \cite{Sahoo2018}.
    
    \textit{Conclusions:} Our generic model calculations suggest that it is feasible to externally functionalize a homogeneous 2D metallic layer by means of structured dielectric environments, thus  creating new plasmonic waveguides using existing experimental techniques and available layered materials. In contrast to previous functionalization concepts, our approach only relies on a passive pre-structured environment, to which the active layer needs to be exposed. Depending on the spatially modulated dielectric contrast in this environment plasmons can be confined within a $10\,$nm scale. At the same time, the described optimal excitation energy window for our proposal renders these devices highly specific to external stimuli allowing for switching or filtering applications.
    The optimization of these new plasmonic devices will thereby be a trade-off between the plasmonic localization and relative brightness which is controlled by the dielectric contrast in the heterogeneous environment.
    
    Furthermore, we highlight that the described induced plasmonic functionality relies on  spatial structuring of the dynamically screened Coulomb interaction within the material. Thus, it can also affect a variety of other many-body properties, including many-body excitations, such as magnons, and many-body instabilities, such as superconductivity or magnetism. The proposed plasmonic waveguide is therefore just one possible example of a more general concept for Coulomb-engineering of many-body properties in metallic layered materials with a variety of further applications.

\begin{acknowledgements}

The authors thank Frank Koppens, Mikhail I. Katsnelson, and Henning Schl{\"o}mer for useful discussions. This work was supported by the US Department of Energy under grant
number DE-FG02-05ER46240. The numerical computations were carried out on the University of Southern California High Performance Supercomputer Cluster. 

\end{acknowledgements}

%%%%%%%%%%%%%%%%%%%%%%%%%%%%%%%%%%%%%%%%%%%%%%%%%%%%%%%%%%%%%%%%%%%%%
%% Start writing some supplementary info here
%%%%%%%%%%%%%%%%%%%%%%%%%%%%%%%%%%%%%%%%%%%%%%%%%%%%%%%%%%%%%%%%%%%%%

\appendix
\section{Background Screening Model}\label{BackScreening}
\begin{figure}[htbp]
 \includegraphics[width=0.8\columnwidth]{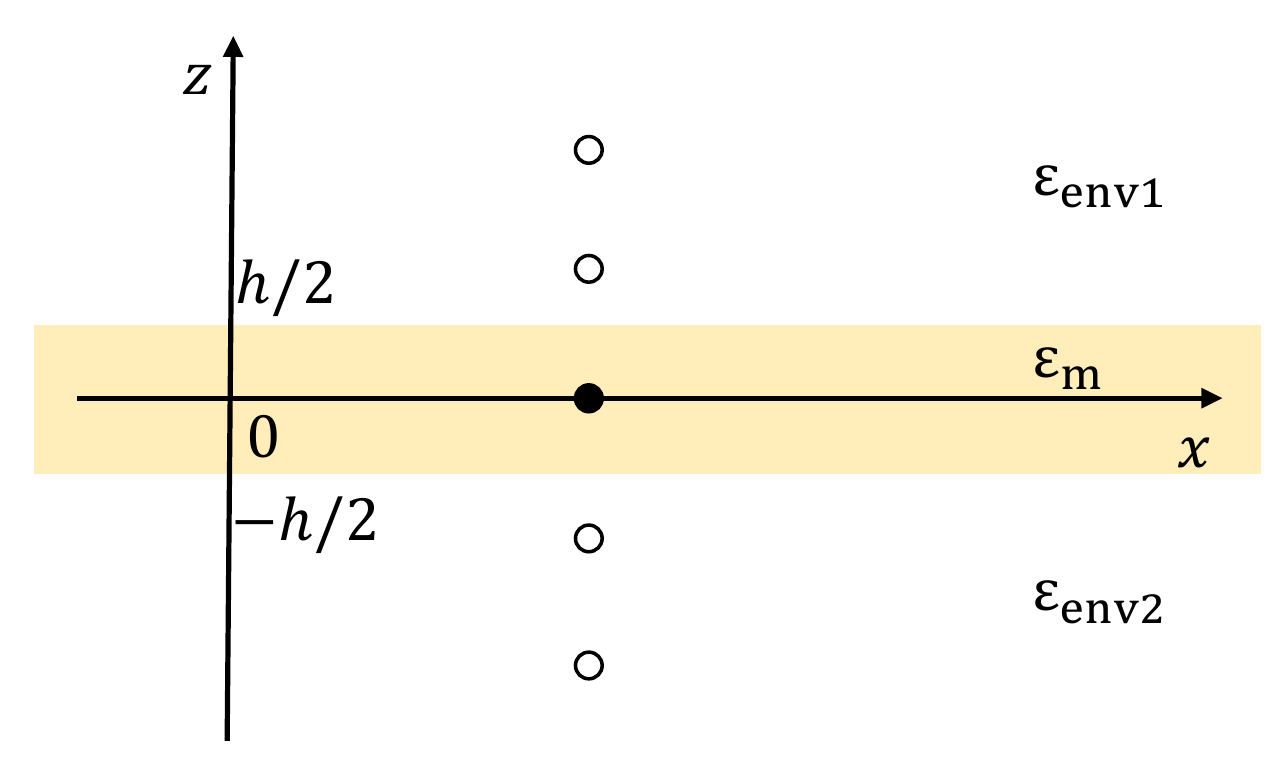}
 \caption{Internal screening in a 2D material along the $z=0$ plane modeled by a dielectric slab of height $h$ with dielectric constant $\varepsilon_m$.}\label{figapp1_screening}
\end{figure}
In vacuum the Coulomb interaction between two point charges $q_1$ and $q_2$ is simply given by $v(r) \propto q_1q_2 / r$, where $r$ is the distance between them. In the case of electrons within a realistic material the effective interaction between them is more complicated due to the polarizable environment \cite{DaJornada2020,Groenewald2016,Rosner2015,Rosner2018}. In layered materials, this internal ``background'' screening can be classically well approximated by a dielectric-slab model. As illustrated in Fig.~\ref{figapp1_screening}, the electrons are supposed to be confined in the center ($z=0$) plane of the dielectric slab, which is defined by the effective height $h$ and the dielectric constant $\varepsilon_\text{m}$. These two parameters are determined from the intrinsic properties of the 2D material and can be calculated from first-principles \cite{Rosner2015,Rosner2016,Groenewald2016}. Here we set $h=5.76$ \AA\ and $\varepsilon_\text{m}=10$ similar to the situation in transition mental dichalcogenide monolayers \cite{Steinke2020}. $\varepsilon_\text{env1}$ and $\varepsilon_\text{env2}$ are the environmental dielectric constants above and below the 2D material.

The ``background'' screened Coulomb interaction $U(r)$ between two electrons in the $z=0$ plane with a separation $r$ can then be determined using an iterated image-charge ansatz \cite{Cho2018,Jackson1999,Kumagai1989}. The image charges (empty circles) generated by the real charge (solid circle) are shown in Fig.~\ref{figapp1_screening}. In principle, there is an infinite number of image charges along the $z$-direction and their charges are determined by $\varepsilon_\text{m}$, $\varepsilon_\text{env1}$, $\varepsilon_\text{env2}$ and $h$. The full ``background'' screened Coulomb interaction $U(r)$ on the $z=0$ plane is given by the sum of all of these contributions. Here, we consider $\varepsilon_\text{env1} = \varepsilon_\text{env2} = \varepsilon_\text{b}$. Then, $U(r)$ is given by
\begin{align}
    U(r) = \frac{e^2}{\varepsilon_\text{m} r} + 2\sum_{n=1}^{\infty} \frac{e^2 \beta_\text{b}^n}{\varepsilon_\text{m} z_n(r)} \label{Vr_IC},
\end{align}
where $\beta_\text{b} = (\varepsilon_\text{m} - \varepsilon_\text{b})/(\varepsilon_\text{m} + \varepsilon_\text{b})$ and $z_n(r) = \sqrt{r^2 + (nh)^2}$. The first term in Eq.~(\ref{Vr_IC}) is the contribution from the real (source) charge, whereas the second term results from the image charges.

For the on-site interaction, i.e. $r=0$, the above formula diverges. To avoid this, we define the ``on-site" interaction at a slightly shifted position with a small separation $\delta$ above the source charge itself. In the numerical calculation we set $\delta=0.85$ \AA, which yields the on-site Coulomb energy to be $2.56\ \text{eV}$ for the unsupported layer. 

\section{Image Charge Ansatz for Spatially Structured Substrates}\label{ICmethod}

\begin{figure}[htbp]
 \includegraphics[width=8.6cm]{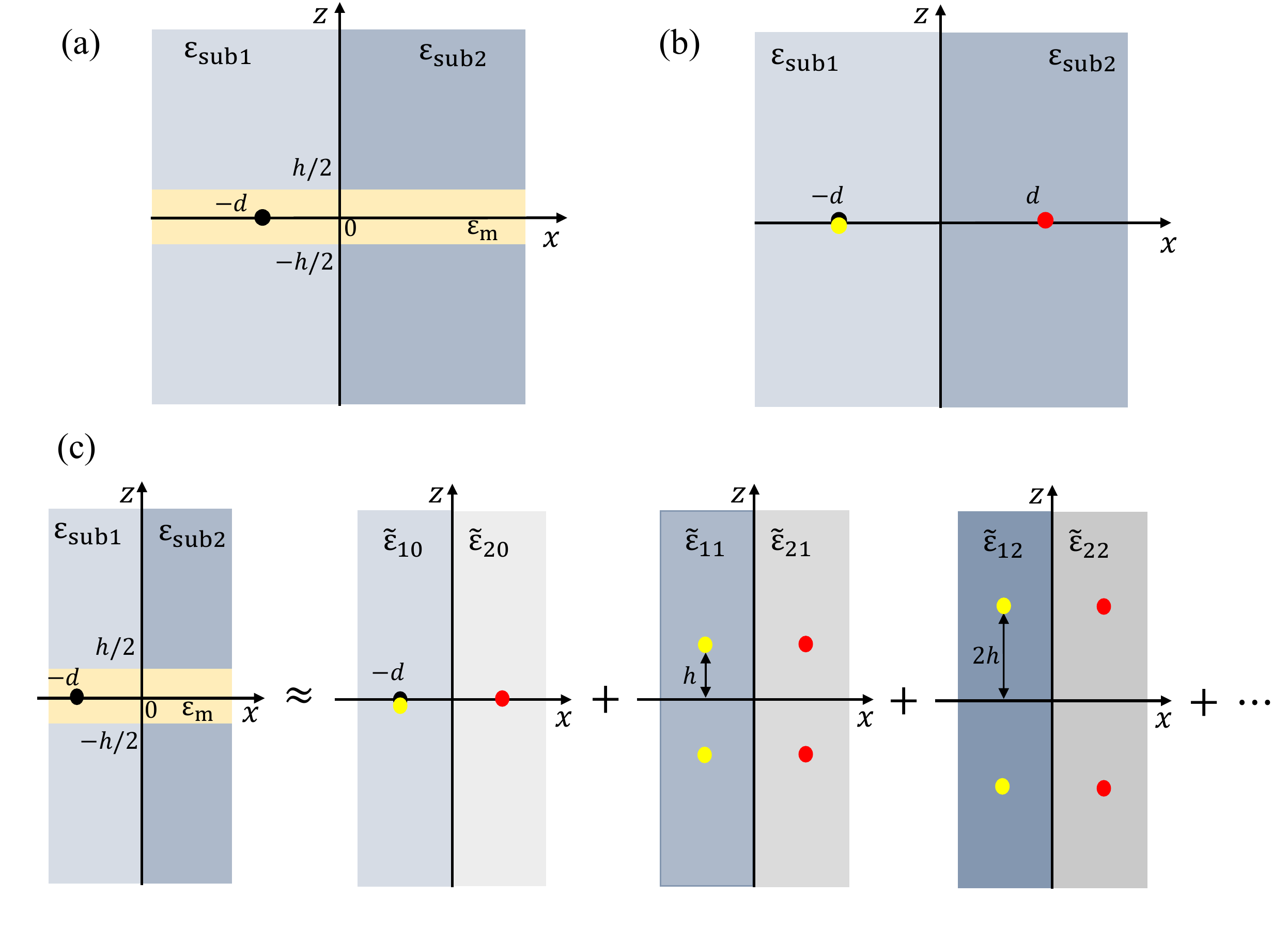}
 \caption{Image charge model of a 2D material embedded in a dielectric structure with a single vertical interface.}\label{figapp2_imagecharge}
\end{figure}

As soon as we additionally introduce vertical dielectric interfaces in the screening environment, the effective Coulomb interaction for $z=0$ can analytically only be approximately described.
The configuration with a single vertical interface is illustrated in Fig.~\ref{figapp2_imagecharge}~(a). In the following we put a real charge $q$ (solid circle) on the $z=0$ plane at a distance of $d$ to the vertical interface. The aim is to approximate the Coulomb potential on the full $z=0$ plane from this point charge.

To this end we combine the analytically known solutions for the homogeneous dielectric slab (two parallel horizontal dielectric interfaces) with the homogeneous single dielectric interface situation depicted in Fig.~\ref{figapp2_imagecharge}~(b). 
In the latter case the analytically correct potential can be constructed using a single additional image charge. In detail, if we place our source charge at $(-d,0)$ [black dot in Fig.~\ref{figapp2_imagecharge}~(b)] we will need an additional image charge at $(d,0)$. For $\vec{r}_x < 0$ this yields
\begin{equation}\label{simple_ic_1}
    U(r) = \frac{e^2}{\varepsilon_{\text{sub}_1}}\left(\frac{1}{r_q} + \frac{\alpha_1}{r_1}\right)
\end{equation}
with $\alpha_1 = \frac{ \varepsilon_{\text{sub}_1}-\varepsilon_{\text{sub}_2}}{\varepsilon_{\text{sub}_1}+\varepsilon_{\text{sub}_2}}$, $r_q = |\vec r + d\hat{x}|$ and $r_1 = |\vec r - d\hat{x}|$. For $\vec{r}_x > 0$ we correspondingly get
\begin{equation}\label{simple_ic_2}
    U(r) = \frac{e^2}{\varepsilon_{\text{sub}_2}} \left(\frac{1}{r_q} + \frac{\alpha_2}{r_2}\right)
\end{equation}
with $\alpha_2=\frac{\varepsilon_{\text{sub}_2}-\varepsilon_{\text{sub}_1}}{\varepsilon_{\text{sub}_1}+\varepsilon_{\text{sub}_1}}$ and $r_2 = r_q$.
This method is discussed in many text books and we refer the interested reader, e.g., to Ref.~\onlinecite{Jackson1999}.

The Coulomb interaction on the $z=0$ plane, before introducing the vertical interface, is already discussed in Appendix~\ref{BackScreening} and given by Eq.~\eqref{Vr_IC}, which we can slightly reformulate to
\begin{align}
    U(r) &= \sum_{n=-\infty}^{\infty} \frac{e^2}{z_n(r) \tilde{\varepsilon}_{\text{b},n}}. \label{Vr_IC_reform}
\end{align}
In this way, we can interpret $U(r)$ as the Coulomb interaction between an electron at $r$ with a series of (electron) point charges positioned at $z_n$ which are each embedded in different homogeneous dielectric backgrounds $\tilde{\varepsilon}_{\text{b},n} = \varepsilon_\text{m} / \beta_\text{b}^{|n|}$.

Now, to model the effect of the additional vertical dielectric interface (between $\varepsilon_\text{sub1}$ and $\varepsilon_\text{sub2}$), we just need to introduce vertical image charges for each of these iterated horizontal image charge as described in Eqs.~(\ref{simple_ic_1}, \ref{simple_ic_1}) and illustrated in Fig.~\ref{figapp2_imagecharge}.
The full approximate interactions can thus be written as a summation of the form
\begin{align}
    U(r) = \sum_n U_n(r),
\end{align}
where each $U_n(r)$ is evaluated as in the simple image charge method introduced before, however, with adjusted parameters. 
Specifically we need to replace $\varepsilon_{\text{sub}_1}$ and $\varepsilon_{\text{sub}_2}$ by $\tilde{\varepsilon}_{1,n}$ and $\tilde{\varepsilon}_{2,n}$ to evaluate $\alpha_{1,n}$ and $\alpha_{2,n}$, and replace $r_q$ by $z_{n,q}(r) = \sqrt{r_q^2 + (nh)^2}$, $r_1$ by $z_{n,1}(r) = \sqrt{r_1^2 + (nh)^2}$ and $r_2$ by $z_{n,2}(r) = \sqrt{r_2^2 + (nh)^2}$. 
Altogether we get
\begin{equation}\label{full_ic}
  U(r) =
  \begin{cases}
    \sum_n \frac{e^2}{\tilde{\varepsilon}_{1,n}} \left( \frac{1}{z_{n,q}(r)} + \frac{\alpha_{1,n}}{z_{n,1}(r)}\right), & \text{if } \vec{r}_x<0, \\
    \sum_n \frac{e^2}{\tilde{\varepsilon}_{2,n}} \left( \frac{2}{z_{n,q}(r)} + \frac{\alpha_{2,n}}{z_{n,2}(r)}\right), & \text{if } \vec{r}_x>0.
  \end{cases} 
\end{equation}

When two parallel vertical interfaces are introduced in the dielectric background, i.e. like the plasmonic waveguide configuration introduced in the main text, we use the same concept. Now, however, we also need to introduce an infinite series of iterated image charges in $x$-direction.

\section{Plasmonic Excitations in Real Space}

We aim to analyze the dielectric function
\begin{align}
    \varepsilon(\omega) = 1 - U \Pi_0(\omega) \label{Eps_RPA}
\end{align}
in real space. To this end we use the Coulomb interaction models from appendix A and B and evaluate the bare polarization $\Pi_0$ of the metallic band within the random phase approximation. 
Since the the active metallic layer itself is translational invariant we can start in momentum space:
\begin{align}
 \Pi_0(q,\omega) = \frac{1}{\Omega_{\text{BZ}}} \sum_{\sigma\vec{k}}\frac{f(\vec{k})-f(\vec{k+q})}{E(\vec{k})-E(\vec{k+q})+\omega+i\gamma}
\end{align}
with $E(\vec{k})$ being the non-interacting single-particle metallic-band dispersion at $\vec{k}$, $f(\vec{k})$ the corresponding Fermi function, and $\gamma = 0.02\,$eV a finite broadening. 
The real-space representation in the atomic basis can then be obtained via an inverse Fourier transformation \cite{VanSchilfgaarde2011},
\begin{align}
    [\mathbf{\Pi}_0(\omega)]_{ab} = \frac{1}{N} \sum_{\vec{q}} \Pi_0(q,\omega)e^{i\vec{q}\cdot(\vec{R}_a-\vec{R}_b)}.
\end{align} 
Here, the vectors $\vec{R}_{a/b}$ are defined on the real-space lattice. This two-step calculation greatly improves the computational efficiency. Finally, we obtain the real-space dielectric function as a matrix in the atomic basis via
\begin{align}
    [\mathbf{\varepsilon}(\omega)]_{ab} = \delta_{ab} - \sum_c {U}(\vec{R}_a-\vec{R}_c) [\mathbf{\Pi}_0(\omega)]_{cb} \label{Eps_real}.
\end{align}
The plasmonic excitations are identified from a spectral decomposition of the dielectric matrix,
\begin{align}
  \mathbf{\varepsilon}(\omega) = \sum_n \varepsilon_n(\omega) \ket{\phi_n(\omega)}\bra{\phi_n(\omega)},
\end{align}
by selecting the ``leading'' dielectric eigenvalue $\varepsilon_\text{max}(\omega)$ which maximizes the electron energy loss spectrum $\text{EELS}(\omega)\propto-\mathrm{Im}\left[1/\varepsilon_n(\omega)\right]$ together with its eigenvector $\ket{\phi_{\text{max}}(\omega)}$ for each frequency. The real-space charge modulation of a plasmon mode at the frequency $\omega_p$ can be obtained from 
\begin{align}
    \ket{\rho_\text{ind}(\omega_p)} = \mathbf{\Pi}_0 (\omega_p) \ket{\phi_{\text{tot}}}
    = \mathbf{\Pi}_0 (\omega_p) \ket{\phi_{\text{max}}(\omega_p)}
\end{align}
\cite{Fahimniya2020}. Here, we interpret the ``leading'' eigenvector $\ket{\phi_{\text{max}}(\omega_p)}$ as the total potential distribution at the plasmon frequency $\omega_p$, since it solves the equation   
\begin{align}
    \varepsilon(\omega) \ket{\phi_{\text{tot}}}
    = \ket{\phi_{\text{ext}}}
\end{align}
when $\ket{\phi_{\text{ext}}}=0$, namely
\begin{align}
    \varepsilon(\omega_p) \ket{\phi_{\text{max}}(\omega_p)}
    = 0 \, \ket{\phi_{\text{max}}(\omega_p)},
\end{align}
i.e. for $\varepsilon_\text{max}(\omega_p) = 0$.
 The product $\mathbf{\Pi}_0 (\omega_p) \ket{\phi_{\text{max}}(\omega_p)}$ then represents the induced charge distribution in the system. The full method has been reported in previous studies \cite{Wang2015,Westerhout2018,Jiang2020}. Finally, we note that by restricting the continuous position coordinates $\vec{r}$ and $\vec{r}'$ to the discretized lattice positions $\vec{R}_a$ and $\vec{R}_b$ we effectively neglect local-field effects \cite{VanSchilfgaarde2011}.

\bibliography{bib_full}

\end{document}